\documentclass[%
 reprint,
 amsmath,amssymb,
 aps,
]{revtex4-2}

\usepackage{graphicx}
\usepackage{dcolumn}
\usepackage{bm}

\usepackage{color}

\newcommand{\beq}{\begin{eqnarray}}
\newcommand{\eeq}{\end{eqnarray}}
\newcommand{\nn}{\nonumber}

\newcommand{\be}{\begin{equation}}
\newcommand{\ee}{\end{equation}}
\newcommand{\vg}{\sqrt{-g}}

\newcommand{\Odr}{\Omega_{dr,0}}
\newcommand{\Ok}{\Omega_{k,0}}
\newcommand{\OL}{\Omega_{\alpha_0,0}}
\newcommand{\Om}{\Omega_{m,0}}
\newcommand{\Ob}{\Omega_{\alpha_1,0}}
\newcommand{\Og}{\Omega_{\alpha_3,0}}
\newcommand{\Or}{\Omega_{r,0}}

\newcommand{\sgn}{\text{sgn}}

\begin{document}

\preprint{APS/123-QED}

\title{Lovelock type brane cosmology}

\author{Amairani Arroyo}
\email{ramairanigtz@outlook.com}
\affiliation{
Departamento de F\'\i sica, Escuela Superior de F\'\i sica y 
Matem\'aticas del Instituto Polit\'ecnico Nacional, \\ Unidad 
Adolfo L\'opez Mateos, Edificio 9, 07738, Ciudad de M\'exico, 
M\'exico
}

\author{Rub\'en Cordero}
\email{rcorderoe@ipn.mx}
\affiliation{
Departamento de F\'\i sica, Escuela Superior de F\'\i sica y 
Matem\'aticas del Instituto Polit\'ecnico Nacional, 
\\ 
Unidad Adolfo L\'opez Mateos, Edificio 9, 07738, Ciudad de 
M\'exico, M\'exico
}

\author{Giovany Cruz}
\email{giocruz@uv.mx}
\affiliation{Facultad de F\'\i sica, Universidad Veracruzana, 
\\
Paseo No. 112, Desarrollo Habitacional Nuevo Xalapa, 91097, 
Xalapa, Veracruz, M\'exico}

\author{Efra\'\i n Rojas}
\email{efrojas@uv.mx}
\affiliation{Facultad de F\'\i sica, Universidad Veracruzana, 
\\
Paseo No. 112, Desarrollo Habitacional Nuevo Xalapa, 91097, 
Xalapa, Veracruz, M\'exico}

\date{\today}

\begin{abstract}
The cosmological implications of the geodetic brane gravity 
model, enhanced by geometrical terms of Gibbons-Hawking-York 
(GHY) type and Gibbons-Hawking-York-Myers type (GHYM), carefully 
constructed as combinations of intrinsic and extrinsic 
curvatures, are examined. All the geometrical terms under 
study belong to a set named Lovelock-type brane models. The 
combined model gives rise 
to a second-order differential equation of motion.  
Under a Friedmann-Robertson-Walker (FRW) geometry defined on 
a $(3+1)$-dimensional worldvolume, together with a perfect 
fluid matter content, the emerging universe of this model 
evolves in a 5-dimensional Minkowski background yielding 
peculiar facts. The resulting Friedmann-type equation is 
written in terms of energy density parameters, where 
fine-tuning is needed to probe interesting cosmological 
processes close to the current data. In this sense, Lovelock-type 
brane models might underlie the cosmic acceleration.
Indeed, we find that these correction terms become significant 
at low energies/late times. 
The model exhibits self-accelerating (non-self-accelerating) behavior 
for the brane expansion, and in the case where the radiation-like 
contribution due to the existence  of the extra dimension vanishes 
its behavior is the same as the Dvali-Gabadadze-Porrati (DGP) 
brane cosmology and its generalization to the Gauss-Bonnet (GB) 
brane gravity. Likewise, Einstein cosmology is recovered when the 
radiation-like contribution fades away along with the odd 
polynomials in brane extrinsic curvature.
\end{abstract}


\maketitle


\section{Introduction}
\label{sec:intro}

The phenomenon of the acceleration of the 
universe~\cite{Riess:1998,Perlmutter:1999}, which goes 
hand by hand with the presence of dark energy, remains 
one of the most intriguing problems of cosmology.
There is no overwhelming observational evidence at all 
supporting the notion of why the number of space-time 
dimensions of our universe be limited to four. 
In spite of that, it is possible to reproduce many features 
of nature in space times with dimension higher than four. 
In this sense, with a view at great scales, to perform 
experiments in four dimensions which reveal their existence, 
it immediately entails the involvement of gravity. On this 
basis, the evidence of a late-time acceleration scheme 
of the universe has motivated an intense research activity 
in the last two decades. It was an unexpected result since 
general relativity with non-relativistic matter produces 
a decelerated expansion which requires the existence of 
the addition of a cosmological constant or other type of 
exotic energy component of the universe. 
The dark energy behavior can be described through the 
introduction of exotic energy and/or modifications to general relativity such as minimally coupled scalar fields like 
quintessence,
~\cite{Wetterich:1987fm,Zlatev:1998tr}, symmetrons~\cite{Kading:2023hdb,Burrage:2018zuj}, $k$-essence fields~\cite{Chiba:2000,Armendariz-Picon:2000,Armendariz-Picon:2001,Melchiorri:2003,Chimento:2003,Chimento:2004,Garriga:1999,Armendariz-Picon:1999,Putter:2007,Gao:2010}, and galileon fields~\cite{Nicolis:2009,Rham:2010,Goon:2011,Goon:2011J}. 
Additionally, $k$-essence models were proposed as a mechanism 
for unifying dark energy and dark matter~\cite{Scherrer:2004}, 
not overlooking the so-called kinetic gravity braiding models, 
which have a relevant cosmological behavior~\cite{Deffayet:2010,Pujolas:2011,Kimura:2011,Maity:2013}. 
Likewise, dark energy can be modelled by modifying the
general relativity action through $F(R)$ 
models~\cite{Capozziello:2002rd,Capozziello:2003gx}, 
scalar-tensor theories~\cite{Amendola:1999qq,Bartolo:1999sq} 
and Gauss-Bonnet dark energy models \cite{Carroll:2004de,Nojiri:2005vv}. Brane world scenarios share this 
common aim and reproduce dark energy dynamics by considering 
the existence of additional dimensions beyond 
four,~\cite{Maartens:2010ar,Stern2023}.

Long ago, Regge and Teiltelboim (RT)~\cite{RT1975} proposed 
that our universe is as an extended object evolving geodesically 
in a Minkowski background spacetime. Based in this simple 
particle/string-inspired assumption, RT's insight was that 
this proposal would serve to unify quantum mechanics with 
General Relativity (GR). In this view, also known as geodetic 
brane gravity (GBG), the gravitational effect of the brane in the bulk 
is neglected. In a like manner, Rubakov introduced the idea that 
the universe could arise as a topological defect~\cite{Rubakov1983}. 
These proposals did not attract much interest due to the lack of a 
strong phenomenological motivation. 
However, these ideas raised the intriguing possibility that 
geometric models play an essential role in the modified gravity 
theories pursuing to describe the accelerated behavior of the 
universe and its dark matter content. 

Geodetic Lovelock brane gravity generalizes RT model and 
extends its scope in some theoretical directions~\cite{Rojas2013}. 
On a technical level, the second-order Lovelock brane invariants 
defined on the world volume swept out by an extended object 
yield second-order equations of motion. This fact is significant since it ensures that there will be no propagation of extra 
degrees of freedom. On physical grounds, the extrinsic curvature correction terms are expected to improve the accelerated 
behavior of the emerging universes in this framework. In this 
regard, there has been significant progress in the analysis 
of the acceleration behavior provided by the named $K$ brane 
action~\cite{Rojas2012,Rojas2014,Rojas2015}, as 
well as in addressing mathematical aspects of the 
theory~\cite{Rojas2013,Rojas2016,Rojas2019,Rojas2025}.

It is accepted in majority that the observed universe 
requires non-baryonic matter to explain many features 
of the evolution of the universe. 
A trait of modern modified theories of gravity focused 
on cosmology is the emergence of an effective exotic energy, also named dark energy, as a companion to given primitive energy density. 
In this sense, as a result of its layout, within the Lovelock-type
brane gravity framework the so-called dark energy is essentially 
a constructed concept arising from a combination of real 
matter and the effects that produce geometrical terms related 
to the shape of the universe. 
In that regard, result attractive to examine both the 
acceleration behavior of these emerging universes, and the effective
energy content they produce. Indeed, in~\cite{Rojas2012,Rojas2024} the cosmological implications of the GBG enhanced with an 
extra term proportional to the extrinsic curvature of the 
brane in the action was considered. The model presents a late 
time self-accelerated expansion of the universe and when 
the radiation-like term is vanishing the model resembles 
the DGP cosmological brane model~\cite{DGP}.

In this paper, within the framework of geodetic 
Lovelock-type brane gravity which 
includes Einstein gravity, DGP, and GB brane 
gravity,~\cite{Charmousis2002,Kofinas2003},
under certain conditions, we investigate 
its joint cosmological implications, 
and examine the dark energy-matter content. 
At first, we highlight the role played by both 
the GHY- and GHYM-type terms in this framework and write
the form that suits us, for our purposes, of the resulting
equation of motion (eom). Then, by imposing a 
Friedmann-Robertson-Walker-Lemaitre (FRWL) geometry on the 
brane, from the eom we find a constant of motion. This
leads us to identify a master equation that allows us to 
find a Friedmann-type equation. Once this is achieved, 
we analyze the evolution of the universe using an effective 
one-dimensional potential and identify a type of fictitious
matter that play the role of dark matter.
Since we have made some progress in this direction 
on the analysis of the $K$-brane action~\cite{Rojas2014}, 
it is natural to explore the cosmological 
implications of the GHYM-type term,~\cite{Myers1987,Davis2003}, 
in this address. In this connection, we will follows 
the same steps developed in~\cite{Rojas2012} to fill this 
gap in the literature on geodetic extended objects.  
Within the geodetic brane cosmology framework, 
a dark radiation-like term enters the game 
due to its relationship with the conserved bulk energy, 
$\omega$, related to the external timelike 
coordinate, $t$. It parametrizes deviation from Einstein 
cosmology~\cite{Davidson1999,Davidson2003}. In like manner, 
in our approach we find a similar term that generalizes 
its role by parametrizing deviations from certain cosmologies, 
such as in unified brane 
cosmology~\cite{Davidson2006}, with the novelty that it 
now includes Gauss-Bonnet brane cosmology in a certain 
limit. It is remarkable that these 
generalized Lovelock terms present a late time 
accelerated  expansion behavior of the 
universe and represent possible physical models for 
dark energy since they are to reproduce many of 
the physical features provided by  
established theories.

The organization of the paper, which is intended to be 
self-contained, is as follows. In Section~\ref{sec2} 
we covariantly formulate the action which describe the 
Lovelock-type brane cosmology. In Section~\ref{sec3}, under 
an FRW geometry on the brane, the reparametrization invariance 
of the model allows us to immediately integrate the equation 
of motion, resulting in an important integration constant, 
$\omega$, which serves as a fingerprint of the extra dimension 
in this setting. In Section~\ref{sec4}, we obtain the general 
Friedmann-type equation of the model and analyze its 
implications for various limits. In this context, we can 
identify a potential energy function that qualitatively 
exhibits the dynamical properties of the universes emerging 
from our approach. We briefly describe the fictitious dark 
energy as a companion to a primitive energy density 
considered. In Section~\ref{sec5}, we summarize our approach 
and outline other interesting issues that could be addressed.

\section{Lovelock-type brane setting}
\label{sec2}

The most interesting action functionals describing $(p+1)$-dim 
surfaces, $m : x^a \rightarrow X^\mu (x^a)$, are those constructed 
out of geometrical scalars using the induced metric $g_{ab}$ 
and the extrinsic curvatures $K_{ab}^i$, as we shall define 
shortly in terms of $X^\mu$ and its derivatives, 
\be 
S[X^\mu] = \int_m d^{p+1}x \,\vg\, L(g_{ab},K_{ab}^i),
\label{action0}
\ee 
where $g = \det (g_{ab})$, and $i = 1,2,\ldots, N - p - 1$. 
Geometrically, the embedding functions $X^\mu (x^a)$ ($\mu = 
0,1,2,\ldots, N-1$) describes the $(p+1)$-dimensional surface 
$m$ parametrized by the coordinates $x^a$ ($a=0,1,2, \ldots, p$). 
The manifold $m$ represents the worldvolume swept out by 
an extended object, $\Sigma$, evolving in a spacetime $\mathcal{M}$.
When constructing an action functional of the form~(\ref{action0})
the  invariance under reparametrizations of $m$ must be kept intact; 
this means that the field variables $X^\mu$ do not appear explicitly in~(\ref{action0}). In most of the models involved in~(\ref{action0}) 
the equations of motion (eom) that arise are of fourth-order in 
derivatives of $X^\mu$.

In a geodesic cosmological scenario, $m$ plays the
role of our universe and is viewed as a $(3+1)$-dimensional 
hypersurface floating in a $(4+1)$-dimensional Minkowski 
background spacetime with metric $\eta_{\mu\nu}$ ($\mu ,\nu 
= 0,1, \ldots, 4$ and $a,b = 0,1,2,3$, and $i=1$).  
The induced metric as well as the extrinsic curvature on $m$ 
are given by $g_{ab} = \eta_{\mu\nu} X^\mu{}_a X^\nu{}_b$ 
and $K_{ab} = - \eta_{\mu\nu} n^\mu D_a X^\nu{}_b$, respectively. 
Here, $X^\mu{}_a = \partial_a X^\mu$ and $n^\mu$ stand for the 
tangent vectors and the normal vector to $m$ defined implicitly 
by $\eta_{\mu\nu} X^\mu{}_a n^\nu = 0$ and $\eta_{\mu\nu} 
n^\mu n^\nu = 1$. Further, the induced metric defines a unique 
torsionless covariant derivative $\nabla_a$ such that 
$\nabla_a g_{bc} = 0$, and $D_a = X^\mu{}_a D_\mu$ is the 
directional derivative along the tangent 
basis,~\cite{Spivak1970,Defo1995}.

In the Lovelock-type brane gravity, the most general action
leading to second-order eom is~\cite{Rojas2012}, 
\be
\begin{aligned}
S[X^\mu] & = \int_m d^4 x\,\sqrt{-g} \left[\alpha_0 
+ \alpha_1 K + \alpha_2 {\cal R} 
\right.
\\
& \left. + \alpha_3 \left( K^3 - 3K K_{ab}K^{ab}
+ 2 K_a{}^b K_b{}^c{}K_c{}^a \right)  \right],
\end{aligned}
\label{action1}
\ee
where $\mathcal{R}$ is the world volume Ricci scalar,
$K=g^{ab}K_{ab}$ is the mean extrinsic curvature, and 
$\alpha_0,\alpha_1, \alpha_2, \alpha_3$ are phenomenological 
parameters with appropriate dimensions. 
Odd polynomials in the extrinsic curvature of the 
Lovelock-type brane invariants have the form of the 
Gibbons-Hawking-York (GHY) and Gibbons-Hawking-York-Myers (GHYM) invariants, respectively, which are seen as counterterms in the case of the presence of bulk Lovelock invariants.

A clever strategy to obtain the eom is based on exploiting the
inherent geometric properties of the conserved stress tensor,
$f^{a\mu}$, associated with the world 
volume~\cite{Noether2000,Rojas2012,Rojas2025}, 
\be
f^{a \mu}_L = 
\sqrt{-g}\left( \alpha_0 J^{ab}_{(0)} + \alpha_1 J^{ab}_{(1)}
+ \alpha_2 J^{ab}_{(2)} + \alpha_3 J^{ab}_{(3)} \right) 
X^\mu{}_b.
\label{fmu1}
\ee
where the $J^{ab}_{(n)}$ are conserved symmetric tensors, 
$\nabla_a J^{ab}_{(n)} = 0$, given by
\be 
\begin{aligned}
J^{ab}_{(0)} =& g^{ab}, 
\\
J^{ab}_{(1)} =& g^{ab} K - K^{ab},
\\
J^{ab}_{(2)} =& g^{ab} \mathcal{R} - 2 \mathcal{R}^{ab}
= -2 G^{ab},
\\
J^{ab}_{(3)} =& g^{ab} (K^3 - 3KK^c{}_dK^d{}_c +
2K^c{}_dK^d{}_eK^e{}_c) \\
&-3 \mathcal{R} K^{ab}
+ 6KK^a{}_cK^{bc} - 6 K^a{}_c K^c{}_d K^{bd}.
\end{aligned}
\label{J0s}
\ee
Here, $\mathcal{R}_{ab}$ is the world volume Ricci tensor
and $G_{ab}$ the associated Einstein tensor. In 
obtaining~(\ref{fmu1}), the identity $J^{ab}_{(n)}
= L_n g^{ab} - n K^a{}_c J^{bc}_{(n-1)}$ satisfied by the tensors
$J^{ab}_{(n)}$ was considered,~\cite{Rojas2012,Rojas2025}.
The brane trajectories and the conditions that must 
hold to maintain the world volume invariance under 
reparametrizations can be obtained from the normal component 
of the covariant conservation law, $n_\mu \nabla_a 
f^{a \mu}_L =0$, and the corresponding tangential component, 
$\partial_b X_\mu \nabla_a f^{a\mu}_L =0$, respectively. 
Indeed, the dynamics of this model is driven by
\be
\label{eom2}
\mathcal{T}^{ab} K_{ab} = 0,
\ee
in a geometrically oriented geodetic type form, or
\be 
\partial_a \left( \sqrt{-g} \mathcal{T}^{ab}\,\partial_b X^\mu 
\right) = 0,
\label{eom1}
\ee
where
\be
\label{Ttensor}
{\cal T}^{ab} = \alpha_0 J^{ab}_{(0)} + \alpha_1 J^{ab}_{(1)}
+ \alpha_2 J^{ab}_{(2)} + \alpha_3 J^{ab}_{(3)},
\ee
while this is complemented by $\nabla_a J^{ab}_{(n)} =0$
which encrypts the invariance under reparametrizations of 
$m$. Notice that the eom,~(\ref{eom2}) or~(\ref{eom1}) is 
of second-order in derivatives of $X^\mu$. 

If an action matter is included, $S_{\text{\tiny m}} 
= \int_m \sqrt{-g} L_{\text{\tiny m}}$ with a matter Lagrangian 
$L_{\text{\tiny m}} (\varphi (x^a), X^\mu)$ localized on the 
brane, the form of the eom~(\ref{eom2}) remains 
practically unchanged since it only receives an extra 
contribution. Certainly, a variational process applied to 
$S_{\textit{\tiny m}}$ yields $\delta S_{\textit{\tiny m}} = 
\int_m \left[ \partial (\sqrt{-g} L_{\text{\tiny m}})/ \partial 
g^{ab} \right] \delta g^{ab}$. After adding this to the variation 
of model~(\ref{action1}) followed by insertion of the 
variation $\delta g^{ab} = - 2 K^{ab} \phi - 2 \nabla^{(a} 
\phi^{b)}$ where $\phi$ and $\phi^a$ denote 
normal and tangential deformation fields, respectively, of 
the world volume (see~\cite{Defo1995,Rojas2025} for 
details), as well as neglecting a surface boundary term, we find 
\be 
\left( \mathcal{T}^{ab} + T^{ab}_{\text{\tiny m}} \right)
K_{ab} = 0,
\label{eom4}
\ee 
or,
\be 
\partial_a \left[ \sqrt{-g} \left( \mathcal{T}^{ab} 
+ T^{ab}_{\text{\tiny m}} \right) \partial_b 
X^\mu \right] = 0,
\label{eom3}
\ee
where $T_{ab}^{\text{\tiny m}} = - (2/ \sqrt{-g}) \partial
(\sqrt{-g} L_{\text{\tiny m}})/\partial g^{ab}$ is the
world volume energy-momentum tensor.

To close this section, notice that we can rewrite the 
eom~(\ref{eom3}) in a challenging fashion. Given that
$J^{ab}_{(2)} = -2 G^{ab}$, the equation of motion~(\ref{eom4})
can be written as
\be 
\label{eom-dark}
\left[ G^{ab} - \frac{1}{2\alpha_2} \left( T^{ab}_{\text{\tiny m}} 
+ \alpha_0 J^{ab}_{(0)} + \alpha_1 J^{ab}_{(1)}
+ \alpha_3 J^{ab}_{(3)}  \right) \right]K_{ab} = 0.
\ee
In view of~(\ref{Ttensor}) and~(\ref{eom3}) we may introduce
a general structure of the form
\be 
\mathsf{T}^{ab} := \alpha_0 J^{ab}_{(0)} + \alpha_1 J^{ab}_{(1)}
+ \alpha_2 J^{ab}_{(2)} + \alpha_3 J^{ab}_{(3)}
+  T^{ab}_{\text{\tiny m}},
\ee
so that~(\ref{eom3}) becomes
\be 
\partial_a \left( \sqrt{-g} \,\mathsf{T}^{ab} \partial_b
X^\mu \right) = 0,
\label{eom5}
\ee
which looks like a wave-like equation. Further, according to 
the traditional Einstein framework, we can 
rewrite~(\ref{eom-dark}) as follows
\be 
\left( G^{ab} - \kappa T^{ab}_{\text{\tiny m}} -
D^{ab} \right) K_{ab} = 0,
\ee
where $D^{ab} := (1/2\alpha_2) (\alpha_0 J^{ab}_{(0)} 
+ \alpha_1 J^{ab}_{(1)} + \alpha_3 J^{ab}_{(3)})$ and
$\kappa:= 1/2\alpha_2$. This expression is equivalent to
\begin{equation}
G^{ab} - \kappa T^{ab}_{\text{\tiny m}} - \tau^{ab} = 0,
\end{equation}
with $\tau^{ab} = D^{ab} + \mathcal{D}^{ab}$, subject to the condition $\mathcal{D}^{ab}K_{ab} = 0$. This structure naturally suggests interpreting $\tau^{ab}$ as an additional contribution to the ordinary matter source $T^{ab}_{\text{\tiny m}}$. In this sense, as discussed
in~\citep{Davidson2001} and~\cite{Paston}, $\tau^{ab}$ can be
understood as an additional matter termed as
\textit{dark matter} or \textit{embedding matter}.
Notice that in our case, such a fictional matter results
in a solely geometric sum of terms.

It is worth pointing out that, given the geometrical origin 
of the Lovelock type brane tensors~(\ref{J0s}), the condition
$\nabla_a D^{ab} = 0$ is fulfilled. This fact would allow us 
to explore certain types of conserved currents.

\section{Lovelock type brane cosmology}
\label{sec3}

By assuming a homogeneous, isotropic and closed universe, 
$m$ can be described by the parametric 
representation
\be
\label{embedding}
x^\mu = X^\mu(x^a) = ( t(\tau),a(\tau), \chi, \theta, \phi),
\ee
where $\tau$ is the proper time measured by an observer at 
rest with respect to the brane. Since $m$ evolves in a 
geodetic form in a 5-dim Minkowski spacetime, 
under~(\ref{embedding}), the ambient spacetime metric can be 
written as $ds_5 ^2 = G_{\mu\nu} dx^\mu dx^\nu = - dt^2 + da^2 + 
a^2 d\Omega_3 ^2$ where $d\Omega_3 ^2 = d\chi ^2 + \sin^2 \chi 
d\theta^2 + \sin^2 \chi \sin^2 \theta d\phi^2$. We introduce 
the lapse function $N := \sqrt{\dot{t}^2 - \dot{a}^2}$, where 
an overdot indicates differentiation with respect to $\tau$. 
Hence, a Friedmann-Robertson-Walker (FRW) line element on $m$
is given by
\be
ds_4 ^2 = g_{ab} dx^a dx^b = - N^2 d\tau^2 + a^2 d\Omega_3 ^2.
\ee
The orthonormal basis describing the world volume is provided 
by four tangent vectors $e^\mu {}_a := X^\mu{}_a$ and the unit 
spacelike normal vector $n^\mu = \frac{1}{N}(\dot{a},\dot{t},0,0,0)$. 
The non-vanishing components of the extrinsic curvature are
\be
K^\tau{}_\tau = \frac{\dot{t}^2}{N^3} \frac{d}{d\tau} 
\left( \frac{\dot{a}}{\dot{t}}\right) 
\quad \text{and} \quad
K^\chi{}_\chi = K^\theta{}_\theta = K^\phi{}_\phi = 
\frac{\dot{t}}{N a}.
\label{Ks}
\ee
These help to compute the non-vanishing components of the 
Lovelock-type brane tensors~(\ref{J0s}),
\begin{widetext}
\be 
\begin{array}{ll}
J^\tau_{(0)\tau} = 1 
& \qquad \qquad 
J^\chi_{(0)\chi} = \, J^\theta_{(0)\theta} = \,J^\phi_{(0)\phi} 
= \,1
\\
J^\tau_{(1)\tau} = \frac{3\dot{t}}{Na}
& \qquad \qquad
J^\chi_{(1)\chi} = \,J^\theta_{(1)\theta} = \,J^\phi_{(1)\phi} 
= \, \frac{\dot{t}}{N^3a} \left[ a\dot{t} \frac{d}{d\tau} 
\left( \frac{\dot{a}}{\dot{t}} \right) + 2 N^2 \right]
\\
J^\tau_{(2)\tau} = \frac{6\dot{t}^2}{N^2a^2}
& \qquad \qquad 
J^\chi_{(2)\chi} = \,J^\theta_{(2)\theta} = \,J^\phi_{(2)\phi} 
= \, \frac{2\dot{t}^2}{N^4a^2} \left[ 2a\dot{t} \frac{d}{d\tau} 
\left( \frac{\dot{a}}{\dot{t}} \right) +  N^2 \right]
\\
J^\tau_{(3)\tau} = \frac{6\dot{t}^3}{N^3a^3}
& \qquad \qquad
J^\chi_{(3)\chi} = \,J^\theta_{(3)\theta} = \,J^\phi_{(3)\phi} 
= \, \frac{2\dot{t}^3}{N^5a^3} \left[ 3 a\dot{t} \frac{d}{d\tau} 
\left( \frac{\dot{a}}{\dot{t}} \right) \right].
\end{array}
\label{J1s}
\ee
\end{widetext}

Based on these points, assuming that the matter on $m$ is 
a perfect fluid, the energy-momentum is
\be
\label{Tab}
T^{ab}_{\text{\tiny m}} = (\rho + P)\eta^a \eta^b + P g^{ab},
\ee
with $P=P(a)$ being the pressure and $\rho = \rho(a)$ the 
energy density, and $\eta^a$ is a timelike unit normal vector 
to $m$ at a fixed time $t$. From~(\ref{Tab}) we extract 
$T^\tau{}_{\text{\tiny m}\tau} = - \rho$ and 
$T^\chi{}_{\text{\tiny m} \,\chi} = T^\theta{}_{\text{\tiny m}
\,\theta} = T^\phi{}_{\text{\tiny m}\,\phi} = P$.

Within the cosmological framework~(\ref{embedding}), 
when $\mu = t$, it produces~(\ref{eom5}) to
become $\partial_\tau (\sqrt{-g}\, \mathsf{T}^{\tau\tau} 
\partial_\tau X^t) = 0$ so that we have one independent 
equation of motion in agreement with the existence of a 
single independent equation provided by~(\ref{eom4}). 
This strategy offers the benefit of integrate this and 
causing the appearance of an important integration 
constant, $\omega$. Indeed, from~(\ref{J1s}) and~(\ref{Tab}) 
we get $\mathsf{T}^{\tau \tau} = - \frac{\alpha_0}{N^2} - 
\frac{\alpha_1}{N^3} \frac{3 \dot{t}}{a} - 
\frac{\alpha_2}{N^4} \frac{6\dot{t}^2}{a^2}
- \frac{\alpha_3}{N^5} \frac{6\dot{t}^3}{a^3} + 
\frac{\rho}{N^2}$, in addition to $g = - N^2 a^6 \sin^4 
\chi \sin^2 \theta$. It follows from these and the only 
single independent equation arising from~(\ref{eom5}) that
\be 
\partial_\tau \left[ - \frac{6 \alpha_3 \dot{t}^4}{N^4}
- \frac{6 \alpha_2 a \dot{t}^3}{N^3} - \frac{3 \alpha_1 a^2 
\dot{t}^2}{N^2} - \frac{(\alpha_0 - \rho)a^3 \dot{t}}{N} 
\right] = 0.
\label{eom6}
\ee
This equation should be accompanied by the integrability
condition $\dot{\rho} + 3 \left( \frac{\dot{a}}{a}
\right) (\rho + P) = 0$ derived from the energy-momentum 
conservation law $\nabla_a T^{ab}_{\text{\tiny m}} = 0$.

A direct integration followed by choosing the cosmic gauge, 
$N=1$, as well as the inclusion of the three main geometries 
by $\dot{t}\to (\dot{a}^2 + k)^{1/2}$ with $k=-1,0,1$, as 
discussed in detail in~\cite{Rojas2012,Rojas2024} 
allows us to find
\be 
\begin{aligned}
- 6\alpha_3\, (\dot{a}^2 + k)^2 - 6\alpha_2 \,a (\dot{a}^2 
+ k)^{3/2} 
\\
- 3 \alpha_1\, a^2 (\dot{a}^2 + k)
- (\alpha_0 - \rho)a^3 (\dot{a}^2 + k)^{1/2} 
:= 6\,\omega,
\end{aligned}
\label{master}
\ee
with $\omega$ being a constant which is related to the conserved
bulk energy conjugate to the embedding time coordinate $t(\tau)$
while the numerical factor has been introduced for convenience.
By rewriting this, we arrive to
\begin{widetext}
\be 
\left( \frac{\dot{a}^2}{a^2} + \frac{k}{a^2}\right)^{1/2}
\left[
\frac{(\alpha_0 - \rho)}{6} 
+ \frac{\alpha_1}{2} \left( \frac{\dot{a}^2}{a^2} 
+ \frac{k}{a^2}\right)^{1/2} + \alpha_2
\left( \frac{\dot{a}^2}{a^2} + \frac{k}{a^2}\right)
+ \alpha_3 \left( \frac{\dot{a}^2}{a^2} 
+ \frac{k}{a^2}\right)^{3/2}
\right]
= - \frac{\omega}{a^4},
\label{master1}
\ee
\end{widetext}
which is a quartic equation for $\left( \dot{a}^2/a^2 + k/a^2 
\right)^{1/2}$. This is a far-reaching equation. On one hand, 
it conveniently furnishes Friedmann-type equations for 
the universes arising in this framework, which allows 
us to explore different possibilities for the cosmological 
expansions. In passing, we mention that such Friedmann 
equations will correspond to first integrals of~(\ref{eom3}). 
On the other hand, this helps to determine dark matter content 
behind these universes, as suggested at the end of 
Section~\ref{sec2}, which will be addressed below. Furthermore,
it is worthwhile to comment that at this level, the
integration constant $\omega$ is of a non-small nature; it can be fine-tuned once the resulting effective energy is determined for comparison with the actual data. In this regard, it can be positive
or negative.

We find it convenient to rewrite~(\ref{master1}) in terms of 
energy density parameters. First, given that we are interested in 
early- and late-time evolution of these universes, we assume 
a total energy density of the form $\rho = \rho_m + \rho_r = 
\rho_{m,0}/a^3 + \rho_{r,0}/a^4$, where $\rho_{m,0}$
and $\rho_{r,0}$ are the current energy densities for matter 
and radiation energy, respectively, as the scale factor is 
fixed at the current time as $a_0=1$. If we introduce the Hubble rate 
$H:= \dot{a}/a$, and the energy density parameters
\be 
\begin{aligned}
\Odr &:= \frac{\omega}{\alpha_2 H_0^3}, & 
\OL  &:= \frac{\alpha_0}{6 \alpha_2 H_0^2},  & 
\Ob  &:= \frac{\alpha_1}{2\alpha_2 H_0},
\\
\Ok  &:= - \frac{k}{H_0^2},  & \,\,
\Om  &:= \frac{\rho_{m,0}}{6\alpha_2 H_0^2}, & 
\Or  &:= \frac{\rho_{r,0}}{6\alpha_2 H_0^2},
\\
\Og  &:= \frac{\alpha_3 H_0}{\alpha_2}, & &
\end{aligned}
\label{densities}
\ee
with $H_0$ being the Hubble constant, then~(\ref{master1}) reads
\begin{widetext}
\be 
\left( \frac{H^2}{H_0^2}
- \frac{\Ok}{a^2} \right)^{1/2} \left[ 
 \left( \frac{H^2}{H_0^2} - \frac{\Ok}{a^2} \right)
+ \left( \OL - \frac{\Om}{a^3} - \frac{\Or}{a^4}\right)
\right]
+ \Ob  \left( \frac{H^2}{H_0^2} - \frac{\Ok}{a^2} 
\right) + \Og  \left( \frac{H^2}{H_0^2}
- \frac{\Ok}{a^2} \right)^2 = - \frac{\Odr}{a^4}.
\label{master2}
\ee
\end{widetext}

A number of remarks are in order. 
Note that the Einstein limit is approached by making $\Odr = 
\Og = \Ob = 0$. Indeed, in such a case, we find the usual 
Friedmann equation in perfect agreement with the standard 
cosmology. Alike, when $\Odr = \Og = 0$,~(\ref{master2}) 
becomes 
\be 
H^2 + \frac{k}{a^2} + \Ob \sqrt{H^2 + \frac{k}{a^2}}
= - \OL + \frac{\Om}{a^3} + \frac{\Or}{a^4},
\ee
which corresponds to the Friedmann-type equation emerging
in the DGP approach~\cite{DGP,Rojas2012}. As discussed 
in~\cite{Rojas2012}, the self-accelerating and 
non-self-accelerating branches are accommodated in the positive 
or negative values of $\Ob$, respectively. 
In a like manner, if only $\Odr = 0$,  it is straightforward 
rewrite~(\ref{master2}) as follows
\begin{widetext}
\be 
 \Ob^2 \left[ 1 + \frac{\Og}{\Ob} \left( \frac{H^2}{H_0^2}
- \frac{\Ok}{a^2}\right) \right]^2 \left( \frac{H^2}{H_0^2}
- \frac{\Ok}{a^2}\right)
= \left[ \frac{H^2}{H_0^2} - \frac{\Ok}{a^2}
- \left( \frac{\Om}{a^3} + \frac{\Or}{a^4} - \OL \right) 
\right]^2.
\ee
\end{widetext}
This equation is clearly in accord with the Friedmann-type 
equation that arise for a braneworld in Gauss-Bonnet gravity in addition of induced gravity, (see (2.13) in~\cite{Kofinas2003}, 
for comparison).

In this spirit, Lovelock type brane cosmology through~(\ref{master2}), unifies these mentioned cosmologies.

\section{Friedmann-like equations}
\label{sec4}

To find the general Friedmann-type equation,
with the aid of
\be 
\label{chi0}
\chi := \left( 
\frac{\dot{a}^2}{a^2} + \frac{k}{a^2}
\right)^{1/2},
\ee
the relationship~(\ref{master2}) can be expressed in the form
\begin{widetext}
\be 
\Og\,\chi^4 + H_0\,\chi^3 + \Ob\,H_0^2\,\chi^2 + \left[
\OL - \left( \frac{\Om}{a^3} + \frac{\Or}{a^4}\right)
\right]H_0^3 \chi + \frac{\Odr\,H_0^4}{a^4} = 0,
\label{quartic}
\ee
\end{widetext}
where we have considered the combination of matter content 
densities, attempting to describe an early radiation era 
and a late-time acceleration epoch. With the dimensionless 
quantities
\begin{widetext}
\be 
\begin{aligned}
u &:= 1 - \frac{8}{3} \Ob \Og,
\\
f(\Omega_I, a) &:= \frac{1}{2} 
\left\lbrace
2 \,\Ob^3 + 27\,u\,\left( \frac{\Odr}{a^4} \right)
+ 9 \,\Ob 
\left[ 
\left( \frac{\Om}{a^3} \right) + \left( \frac{\Or}{a^4} \right)
- \OL \right]
\right.
\\
& 
\qquad \qquad \qquad \qquad \qquad \qquad \qquad \qquad \qquad
\left.+ \,27 \,\Og \left[ \left( \frac{\Om}{a^3} \right)
+ \left( \frac{\Or}{a^4} \right) - \OL \right]^2
\right\rbrace,
\\
g(\Omega_I,a) &:= \left( \Ob^2 - 3\OL \right) + 12\,\Og 
\left( \frac{\Odr}{a^4} \right)
+ 3 \left[ \left( \frac{\Om}{a^3} \right) + \left( \frac{\Or}{a^4}
\right) \right],
\\
h(\Omega_I,a) &:= 
\left\lbrace
f (\Omega_I,a) 
\left[
1 + 
\sqrt{1 - \frac{[g(\Omega_I,a)]^3}{[f(\Omega_I,a)]^2}}
\right]
\right\rbrace^{1/3},
\\
s(\Omega_I,a) &:= 1 - 4\,\Ob\Og - 8 \,\Og^2
\left[
\left( \frac{\Om}{a^3} \right) + \left( \frac{\Or}{a^4}\right)
- \OL \right],
\end{aligned}
\label{structures}
\ee
\end{widetext}
it follows that 
\begin{widetext}
\be 
\begin{aligned}
\chi &= - \frac{H_0}{4|\Og|} 
\left\lbrace
\sgn (\Og) \pm \sqrt{u + \frac{4}{3} 
\left[ 
\frac{g(\Omega_I,a)}{h(\Omega_I,a)} + h(\Omega_I,a)
\right]
\Og}
\right.
\\
&\qquad \qquad \qquad \qquad \qquad  \quad
\pm \left. \sqrt{
2\,u - \frac{4}{3}
\left[ 
\frac{g(\Omega_I,a)}{h(\Omega_I,a)} 
+ h(\Omega_I,a)
\right]
\Og + 
\frac{\sgn (\Og)\,2\, s(\Omega_I,a)}{\sqrt{u + \frac{4}{3} 
\left[ 
\frac{g(\Omega_I,a)}{h(\Omega_I,a)} + h(\Omega_I,a)
\right]
\Og}
}
}
\right\rbrace,
\end{aligned}
\label{solutions}
\ee
\end{widetext}
represents a whole family of solutions where $\Omega_I$ 
stands for the energy density parameters~(\ref{densities}), 
and $\sgn (\Og) = 
\begin{cases}
\,\,\,\,1, & \quad \text{if}\,\, \Og > 0,
\\
-1, & \quad \text{if}\,\, \Og < 0.
\end{cases}$.
The solutions~(\ref{solutions}) are extremely involved 
so that their real values depend greatly on the values 
of $\Omega_I$ as dictated by the discriminant 
of~(\ref{quartic}),~\cite{Prodanov2021}.
Furthermore, when squaring~(\ref{chi0}), it takes the form
\be
H^2 + \frac{k}{a^2} = \chi^2,
\label{Friedmann0}
\ee
with $\chi(a;\Omega_I)$ being one of the roots~(\ref{solutions}). 
We thus find a set of four Friedmann-like equations 
that, classically, capture physical information about 
the dynamical evolution for this type of universes.

It is clear that by evaluating at the present moment, relationship~(\ref{master2}) reduces to
\be 
\begin{aligned}
\left( 1 - \Ok \right)^{1/2} \left(
1 - \Ok + \OL - \Om - \Or \right) 
\\
+ \,\Ob \left( 1- \Ok \right) + \Og \left( 1 - \Ok \right)^2
= - \Odr.
\end{aligned}
\label{normalization}
\ee
This represents a generalized normalization condition 
that is useful for both performing numerical analysis 
and identifying special limits.

On physical grounds, the real roots of the quartic 
equation we are seeking must satisfy certain conditions
determined by the discriminant of~(\ref{quartic}). Such 
values depend strongly on the values of the parameters 
$\Omega_I$,~\cite{Prodanov2021}.

\subsection{Potential energy functions}
 
By knowing the solutions $\chi$, and by fine-tuning the model 
parameters, we can learn generic features of the dynamical 
behavior of many self-contained universes. Indeed, by rewriting~(\ref{Friedmann0}) in terms of the energy 
densities~(\ref{densities}) we have $\dot{a}^2 - H_0^2 \Ok - 
a^2 \,\chi^2 = 0$. In this fashion, we face an equivalent 
1-dimensional non-relativistic mechanical problem with a 
vanishing total mechanical energy, and where 
$U (\Omega_I;a) := H_0^2 \left( - \Ok - a^2\chi^2/H_0^2\ 
\right)$ represents effective potential energy functions, 
parametrized  by the $\Omega_I$. These can be read off 
immediately
\begin{widetext}
\be 
\begin{aligned}
\frac{U}{H_0^2} &= - \Ok - \frac{a^2}{16\Og^2}
\left\lbrace
\sgn (\Og) \pm \sqrt{u + \frac{4}{3} 
\left[ 
\frac{g(\Omega_I,a)}{h(\Omega_I,a)} + h(\Omega_I,a)
\right]
\Og}
\right.
\\
&\qquad \qquad \qquad \qquad \qquad \qquad \quad \pm \left. 
\sqrt{
2\,u - \frac{4}{3}\left[ \frac{g(\Omega_I,a)}{h(\Omega_I,a)} 
+ h(\Omega_I,a)\right]\Og 
+ \frac{\sgn (\Og)\,2\, s(\Omega_I,a)}{\sqrt{u 
+ \frac{4}{3} 
\left[ 
\frac{g(\Omega_I,a)}{h(\Omega_I,a)} + h(\Omega_I,a)
\right]
\Og}
}
}
\right\rbrace^2.
\end{aligned}
\label{Us}
\ee
\end{widetext}
As already mentioned, the setup reproducing some current 
cosmological behaviors based on recent observational data 
could be a Lovelock-type brane with energy densities 
$\Omega_I$, which need to be properly fine-tuned as we 
shall see shortly.

\subsection{Potential energy functions for special cases}

To uncover particular features of the model with the idea 
of highlighting the role that GHYM-type term plays in 
development, in a scenario without a cosmological constant 
on the brane, we will address some illustrative reductions 
by turning off certain parameters.

\subsubsection{$\OL = 0$, $\Om = 0$, $\Or$, and $\Odr = 0$}

This is a merely geometrical model. In this case the master equation~(\ref{quartic}) reads
\be 
\Og \chi^4 +  H_0 \chi^3 + H_0^2 \Ob \chi^2 = 0.
\ee
In turn, this takes the form of a biquadratic equation
whose solutions are easily found and given by
\beq 
\chi_1 &:=& \frac{-H_0}{2\,\Og} \left( 1 - \sqrt{1 - 4
\Ob\,\Og}
\right),
\label{chi1}
\\
\chi_2 &:=& \frac{-H_0}{2\,\Og} \left( 1 + \sqrt{1 - 4
\Ob\,\Og}
\right).
\label{chi2}
\eeq
Additionally, we get $\chi_{3,4} = 0$. It is clear that, 
depending on the values chosen for 
$\Og$ and $\Ob$ we shall have physical or non-physical 
solutions. As discussed above, by squaring and rearranging 
these solutions, in addition to considering~(\ref{Friedmann0}), 
it follows a pair of Friedmann equations
\be 
\begin{aligned}
H^2 + \frac{k}{a^2} &= H_0^2 \left( \frac{1 - \sqrt{1 - 4\,
\Ob \Og}}{2\,\Og}\right)^2,
\\
H^2 + \frac{k}{a^2} &= H_0^2 \left( \frac{1 + \sqrt{1 - 4\,
\Ob \Og}}{2\,\Og}\right)^2.
\end{aligned}
\ee
Alike, we readily obtain the effective energy potential 
functions
\be
\begin{aligned} 
\frac{U_1}{H_0^2} &= - \Ok - a^2 \left( \frac{1 - \sqrt{1 - 
4 \Og \Ob}}{2\Og} \right)^2,
\\ 
\frac{U_2}{H_0^2} &= - \Ok - a^2 \left( \frac{ 1 + \sqrt{1 - 
4 \Og \Ob}}{2\Og} \right)^2. 
\end{aligned}
\label{Friedmanns2}
\ee
These correspond to distinct branches of the effective 
potential that emerge in this particular scenario. Based 
in the Einstein cosmology, and with support 
of~(\ref{Friedmann0}), the simplest conventional FRW 
equation $\dot{a}^2 + k = (\Lambda/3)a^2$ is retrieved with 
the effective positive cosmological constants 
\be 
\begin{aligned}
\Lambda_{1\,\text{\tiny eff }} &:= 3 H_0^2 
\left( \frac{1 - \sqrt{1 - 4 \Og \Ob}}{2\Og} \right)^2, 
\\
\Lambda_{2\,\text{\tiny eff }} &:= 3H_0^2
\left( \frac{ 1 + \sqrt{1 - 4 \Og \Ob}}{2\Og} \right)^2.
\end{aligned}
\ee
In both cases, the acceleration is driven merely by
geometric effects.
As a matter of fact, the branch $U_2$ leads to 
a significantly faster expansion compared to the branch $U_1$. 
This distinction implies that, although both solutions enter 
an accelerated phase, the expansion rate can vary markedly 
depending on the selected branch.

\begin{figure}[h]
\centering
\includegraphics[width=1\columnwidth]{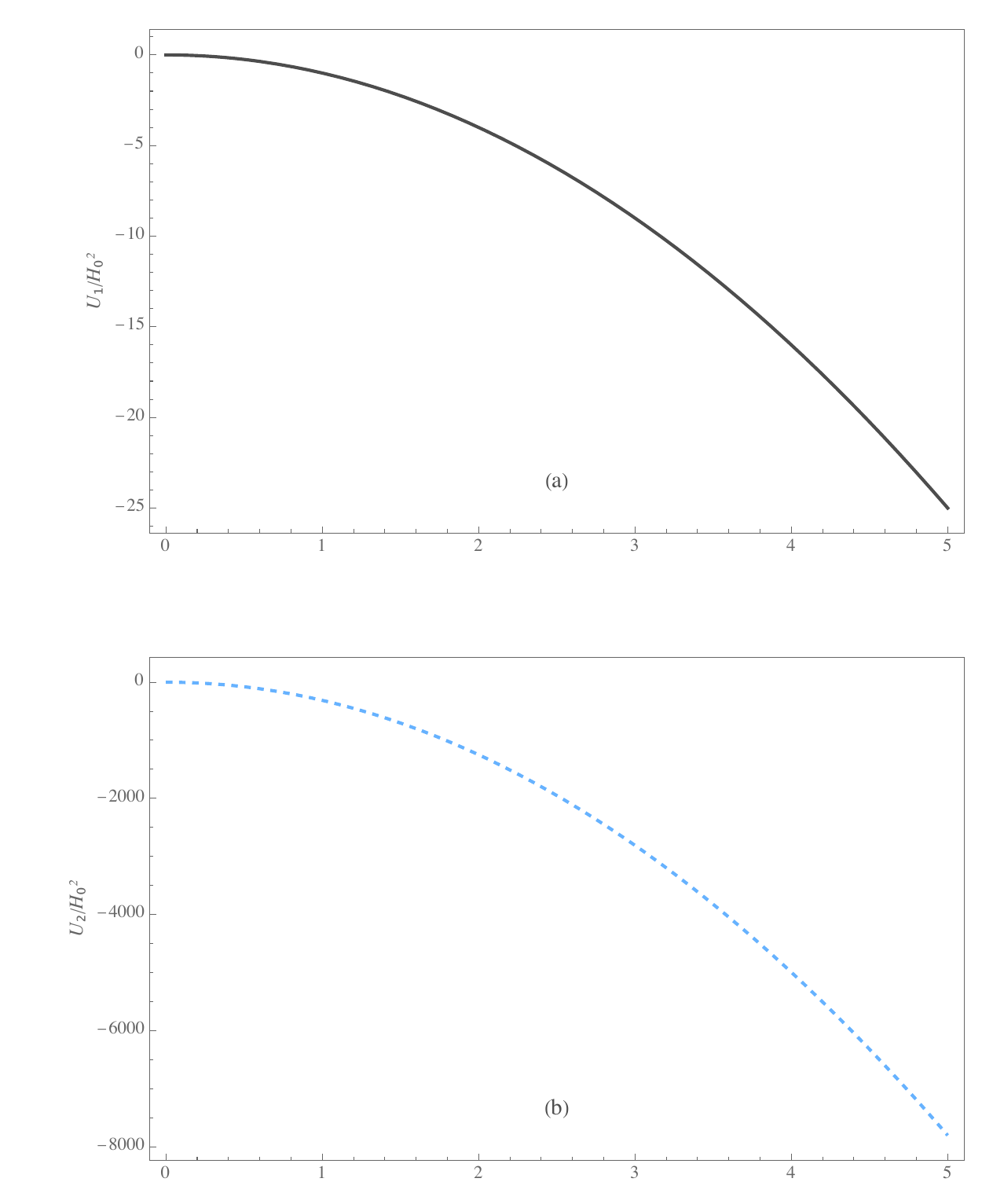}
\caption{Here the parameter values are: $\Ok = 0 $, 
$\Ob = -1.06 $, and $\Og = 0.06$.
} 
\label{fig1}
\end{figure}

To better understand their physical implications, we now 
compare their behavior as functions of the scale factor $a$ 
in Figure~\ref{fig1}, where both branches are plotted for a 
representative choice of parameters.
This case closely resembles the expected behavior of the 
universe in the far future according to the standard 
$\Lambda$CDM model, in which the matter density becomes 
negligible and the cosmological constant entirely governs 
the expansion dynamics. In this regime, the evolution is 
controlled by a single component whose energy density remains 
nearly constant over time, leading to a phase of sustained 
accelerated expansion. 

We can further develop~(\ref{chi1}) and~(\ref{chi2}) as follows.
If $|4\Og\,\Ob| < 1$, then we have the approximated solutions
$\chi_1 \simeq - H_0\,\Ob$ and $\chi_2 \simeq - H_0 \left( \Ob 
- \frac{1}{\Og} \right)$.
These expansions are also reflected in the expressions for the 
potential functions~(\ref{Friedmanns2}) and the form for the
effective cosmological constants. Certainly, these expansions 
yield $\Lambda_{1\,\text{\tiny eff }} \simeq 3 H_0^2 \Ob^2$ and 
$\Lambda_{2\,\text{\tiny eff }} \simeq 3 H_0^2 \left( \Ob - 
\frac{1}{\Og} \right)^2$, which supports the previous description
bearing in mind the corresponding normalization condition provided
by~(\ref{normalization}).

\subsubsection{$\OL = 0$, and $\Odr = 0$}

\begin{figure}[h]
  \centering
  \includegraphics[width=1\columnwidth]{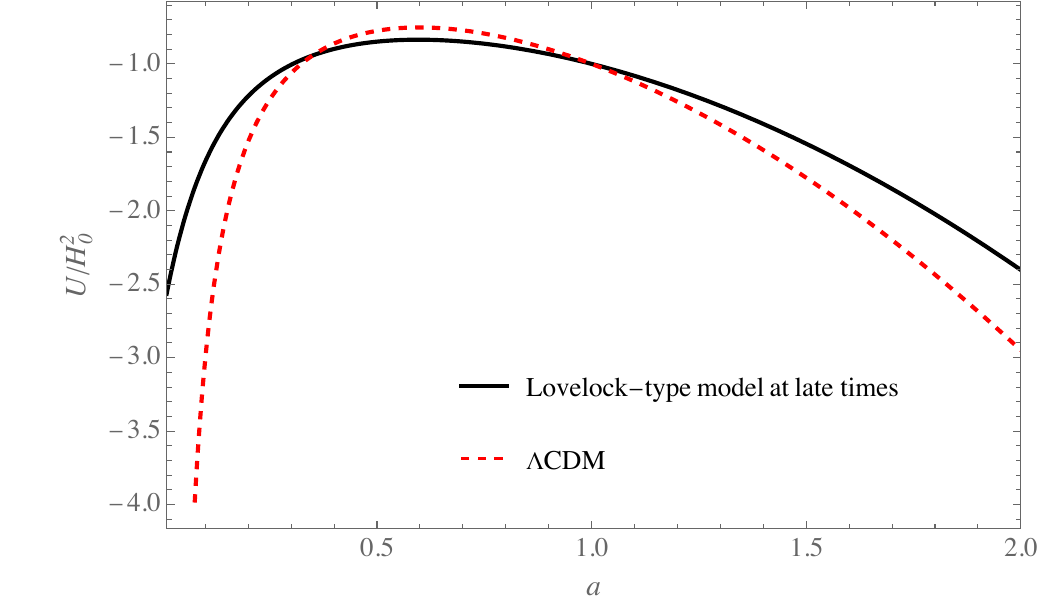}
  \caption{Comparison between~\eqref{ep2} and the potential 
  of the $\Lambda$CDM model. We used the parameter 
  values: $\Omega_{m,0} = 0.3$, $\Omega_{k,0} = 0$, 
  $\Omega_{\alpha_1,0} = -0.76$, and 
  $\Omega_{\alpha_3,0} = 0.06$.} 
  \label{fig2}
\end{figure}

This case allows us to analyze two ordinary cosmological 
epochs of the Universe. If the universe is matter- and 
radiation-dominated then~(\ref{quartic}) takes the form
\be 
\Og\,\chi^3 + H_0 \chi^2 + \Ob\,H_0^2\,\chi - \left(
\frac{\Om}{a^3} + \frac{\Or}{a^4} \right)H_0^3 = 0,
\ee
where one of the solutions $\chi$ vanishes identically.
By virtue of the techniques for solving cubic equations,
we shall have one physical root or three physical ones,
respectively. For the sake of illustration, we confine
ourselves to discuss the case of having one real solution. 
Defining
\beq 
G &:=& 1 - 3\,\Ob\Og,
\label{G}
\\
F(\Omega_I,a) &:=& \frac{1}{2} \left[ 2 - 9 \,\Ob \Og 
\right.
\nn
\\
& & \left. \qquad \qquad - 
27 \,\Og^2 \left( \frac{\Om}{a^3} + \frac{\Or}{a^4}\right)
\right],
\\
\mathsf{H}(\Omega_I,a) &:=& \left\lbrace
F(\Omega_I,a) \left[
1 - \sqrt{1 - \frac{G^3}{[F(\Omega_I,a)]^2}}
\right]
\right\rbrace^{1/3},
\eeq
we get
\be 
\chi = - \frac{H_0}{3 \Og} \left[ 
1 + \mathsf{H}(\Omega_I , a)
+ \frac{G}{\mathsf{H}(\Omega_I , a)}
\right].
\label{solution2}
\ee
It is straightforward to determine the associated effective
potential $U(\Omega_I,a)$
\be 
\frac{U}{H_0^2} = - \Ok - \frac{a^2}{9 \Og^2}
\left[ 
1 + \mathsf{H}(\Omega_I,a) + \frac{G}{\mathsf{H}(\Omega_I,a)}
\right]^2.
\label{ep2}
\ee
To gain intuition about its behavior, on the one hand, we plot 
this in figure~\ref{fig2} to qualitatively compare a Lovelock-type
brane model with the standard $\Lambda$CDM scenario.

As illustrated in figure~\ref{fig2}, both models exhibit a 
qualitatively similar evolution of the potential, particularly 
near the transition to cosmic acceleration. However, our model 
shows a smoother onset of the accelerating phase, 
as a result of a more gradual departure from decelerated 
expansion. Note that, for the chosen parameters the inflection 
point of the potential—interpreted as the moment of acceleration 
onset—occurs together 
in both models. This coincidence, while not intrinsic to 
the modified scenario, offers a natural benchmark for 
comparing the dynamical implications of both frameworks 
under similar conditions.

\begin{figure}[h]
\includegraphics[width=1\columnwidth]{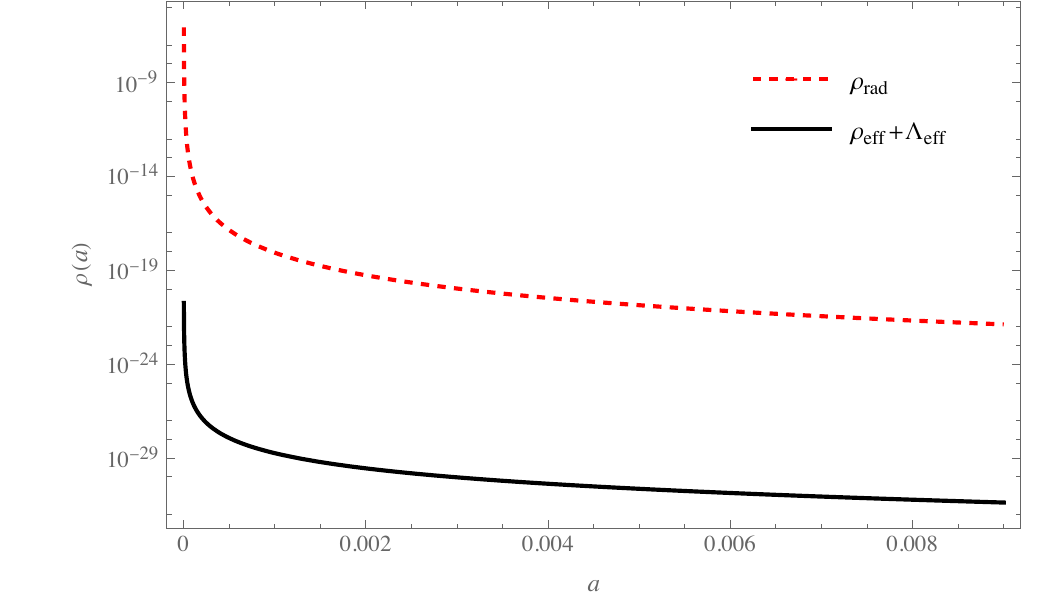}
\caption{The parameters values are $ H_0 = 67.4 \text{km} 
\cdot\text{s}^{-1}\cdot \text{Mpc}^{-1}$, $\Or = 
9\times 10^{-5}$, $\Omega_{k,0} = 0$, $\Ob 
= -1.05$, and $\Og = 0.06$. The vertical axis is shown on 
a logarithmic scale to allow both curves — radiation and 
the dark component — to be displayed together despite the 
large difference in their magnitudes.
}
\label{fig3}
\end{figure}

On the other hand, if we give priority to the inflationary 
era, the energy density is the radiation-dominated one so that 
$\rho= \rho_r/a^4$ and the quartic equation~(\ref{quartic}) 
becomes more analytically manageable by turning off $\Om$.
The solution is provided by~(\ref{solution2}) with $\Om=0$.
In this instance the solution~(\ref{solution2}) 
allows us directly to perform expansions over values 
of $a$. Under this relaxation we have the approximated 
solution around $a=0$
\be 
\begin{aligned}
\chi & \simeq \frac{H_0}{3 \Og}
\left[
\left( \frac{\Omega_{r,\text{\tiny eff}}}{a^4}\right)^{1/3}
- 1 + G \left( \frac{a^4}{\Omega_{r,\text{\tiny eff}}}\right)^{1/3}
\right.
\\
& \left. - \left( G - \frac{1}{3} \right) \left( \frac{a^4}{\Omega_{r,\text{\tiny eff}}}\right)^{2/3}
\right],
\end{aligned}
\label{eq46}
\ee
where $G$ is defined in~(\ref{G}) and we have introduced 
\be 
\Omega_{r, \text{\tiny eff}} := 27 \,\Og^2 \Or.
\label{Omega-r-eff}
\ee 
On squaring~(\ref{eq46}) and considering~(\ref{Friedmann0}), 
followed by an appropriate rearrangement we find the 
Friedmann-type equation
\be
\begin{aligned}
& H^2 + \frac{k}{a^2} = 
\frac{1}{3} \left\lbrace \frac{H_0^2 (1 -  2 \Ob 
\Og)}{\Og^2} 
\right.
\\
& \left. + \frac{H_0^2}{3\Og^2} \left[ \left( 
\frac{\Omega_{r, \text{\tiny eff}}}{a^4} \right)^{2/3} 
- 2 \left( 
\frac{\Omega_{r, \text{\tiny eff}}}{a^4} \right)^{1/3}
\right. \right.
\\
& \left. \left. - 4 \left( G - \frac{1}{6} \right) 
\left( 
\frac{a^4}{\Omega_{r, \text{\tiny eff}}} \right)^{1/3}
\right]
\right\rbrace.
\end{aligned}
\label{eq61}
\ee
Grounded on the traditional form of the Friedmann equation, 
$\dot{a}^2 + k = \frac{1}{3} \Lambda\,a^2 + \frac{1}{3}
\rho \,a^2$, we can immediately read off 
effective parameters. On the one hand, an effective 
cosmological constant
\be 
\Lambda_{\text{\tiny eff}} : = \frac{H_0^2}{\Og^2} (1 -  
2\Ob \Og),
\label{Leff1}
\ee
and an effective energy density
\be
\begin{aligned}
\rho_{\text{\tiny eff}} & = \frac{H_0^2}{3\Og^2}
 \left[ 
  \left( 
\frac{\Omega_{r, \text{\tiny eff}}}{a^4} \right)^{2/3}  
- 2  \left( 
\frac{\Omega_{r, \text{\tiny eff}}}{a^4} \right)^{1/3}  
\right.
\\
& \left. - 4 \left( G - \frac{1}{6} \right) 
 \left( 
\frac{a^4}{\Omega_{r, \text{\tiny eff}}} \right)^{2/3}  
\right],
\end{aligned}
\label{rho-eff}
\ee
where $\Omega_{r,\text{\tiny eff}}$ is defined 
in~(\ref{Omega-r-eff}). These findings show, on the one hand, 
the strong influence of the cubic extrinsic curvature term 
at very early times through $\rho_{\text{\tiny eff}}$, causing 
the universe to expand fast in an unconventional way, 
similar to named Gauss-Bonnet regime in brane cosmology~\cite{Charmousis2002,Kofinas2003}. On the other 
hand, $\Lambda_{\text{\tiny eff}}$ is an entirely 
geometrical effect provided by $\Ob$ and $\Og$.
The figure \ref{fig3} illustrates these effects, providing a qualitative comparison between radiation and the effective density within this model. Unlike in $\Lambda$CDM, where radiation decreases as $1/a^4$ and dominates the early stages of the universe, here the effective density is attenuated and remains several orders of magnitude below the standard radiation.

In the referenced approximation, focusing on the early 
universe, from~(\ref{eq61}) one also finds an effective 
potential function
\be 
\frac{U}{H_0^2} = - \Ok - \frac{a^2}{3\,H_0^2}\left[
\Lambda_{\text{\tiny eff}} + \rho_{\text{\tiny eff}}(a)
\right],
\label{eq50}
\ee
where $\Lambda_{\text{\tiny eff}}$ and $\rho_{\text{\tiny eff}}$
are given by~(\ref{Leff1}) and~(\ref{rho-eff}), respectively.
\begin{figure}[h]
\centering
  \includegraphics[width=1\columnwidth]{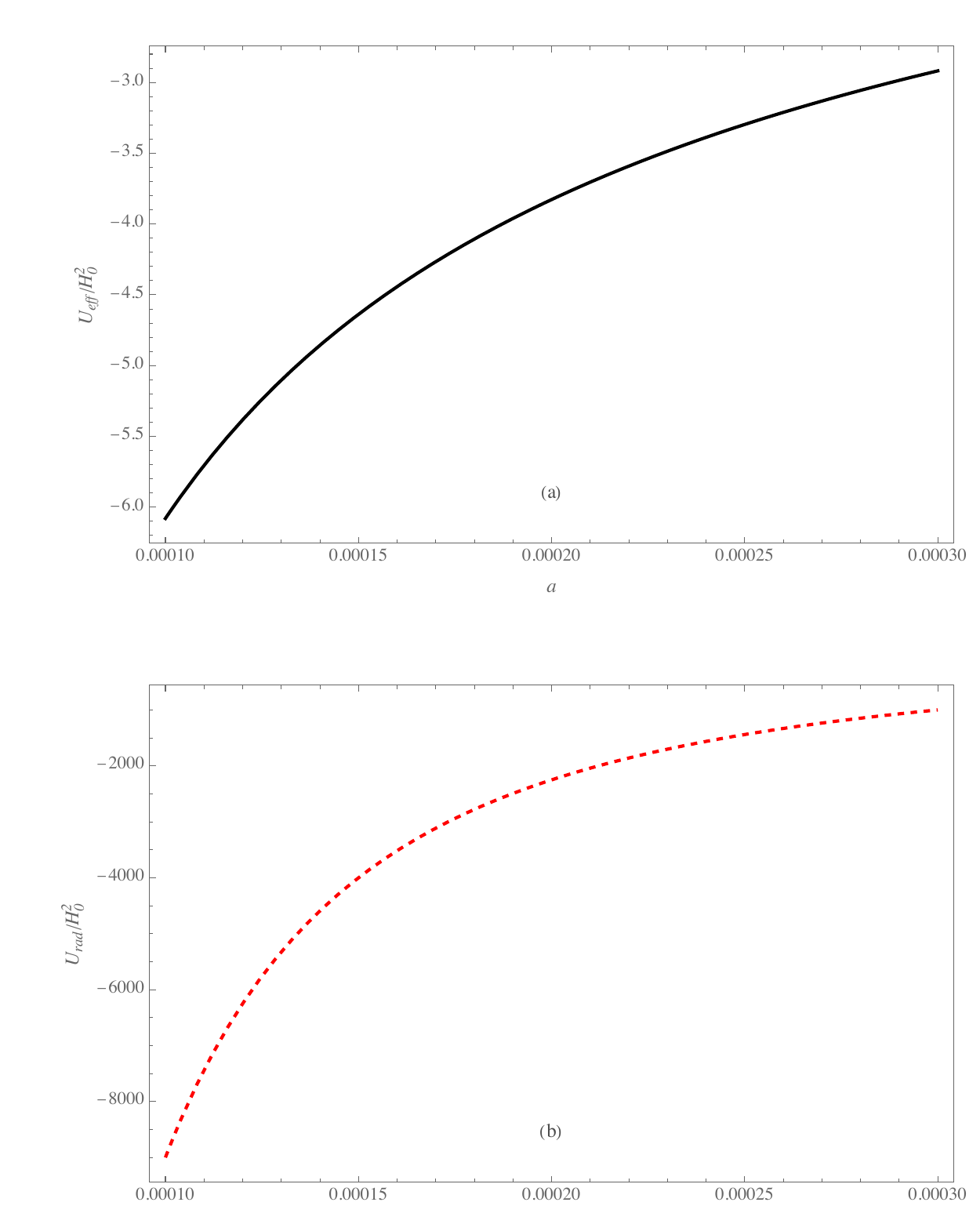}
  \caption{The parameter values are: $ H_0 = 67.4  
  \text{km} \cdot\text{s}^{-1}\cdot \text{Mpc}^{-1} $, 
  $\Or = 9\times 10^{-5}$, $\Ok = 0$, $ \Ob = -1.05 $, 
  and $\Og = 0.06$.} 
  \label{fig3_1}
\end{figure}
In figure~\ref{fig3_1} we depicted the behavior of the 
effective potential~(\ref{eq50}) incorporating Lovelock-type 
terms (black line), compared with the standard radiation 
potential $U_{\text{rad}}(a)$ of the $\Lambda$CDM model 
(red dashed line), at early stages of the universe. Both 
potentials are normalized by $H_0^2$.
Radiation does not appear as an independent term, but is instead included effectively within the structure of the potential, together with other geometric contributions. 
Despite this, $U_{\mathrm{rad}}$ exhibits a much larger negative magnitude than 
$U_{\mathrm{eff}}$. In both cases the absolute value decreases as the universe expands. This trend suggests that the effective component—including radiation—is attenuated or partially compensated by other contributions, thereby softening its dynamic impact during the initial cosmic evolution. To end this discussion, this comparison reveals that although our model incorporates radiation, it is \emph{masked or diluted} within a more complex dynamical structure, preventing it from dominating as clearly as in the pure radiation potential of $\Lambda$CDM.

\subsubsection{$\OL = 0$, and $\Odr \neq 0$}

In this particular instance when $\Odr$ enters the game, 
and intending to focus on a radiation-dominated era, the 
quartic equation~(\ref{quartic}), becomes
\be 
\Og \chi^4 + H_0 \chi^3 + H_0^2\, \Ob\chi^2 - \frac{H_0^3 \Or}{a^4} 
\chi + \frac{H_0^4 \Odr}{a^4} = 0.
\ee
The solutions are quite involved and provided by~(\ref{solutions}).
Given these facts, structures~(\ref{structures}) reduce to
\beq 
u &=& 1 - \frac{8}{3}\Ob \Og,
\nn
\\
f(\Omega_I,a) &=& \frac{1}{2}
\left\lbrace
2\,\Og^3 + 27\,u\,\left( \frac{\Odr}{a^4} \right)
+ 9\,\Ob \left( \frac{\Or}{a^4} \right) 
\right.
\nn
\\
&+& \left. 27\,\Og \left( \frac{\Or}{a^4} \right)^2
\right\rbrace,
\nn
\\
g(\Omega_I,a) &=& \Ob^2 + 12\,\Og \left( \frac{\Odr}{a^4}
\right) + 3 \left( \frac{\Or}{a^4} \right),
\nn
\\
h(\Omega_I,a) &=& \left\lbrace
f \left[ 1 + \sqrt{1 - \frac{g^3}{f^2}} \right]
\right\rbrace^{1/3},
\nn
\\
s (\Omega_I,a) &=& 1 - 4\,\Ob \Og - 8\,\Og^2
\left( \frac{\Or}{a^4} \right).
\nn
\eeq
Likewise, the energy potential functions are provided 
by~(\ref{Us}) taking into considerations the previous 
relationships. 

\begin{figure}[h!]
\includegraphics[width=1\columnwidth]{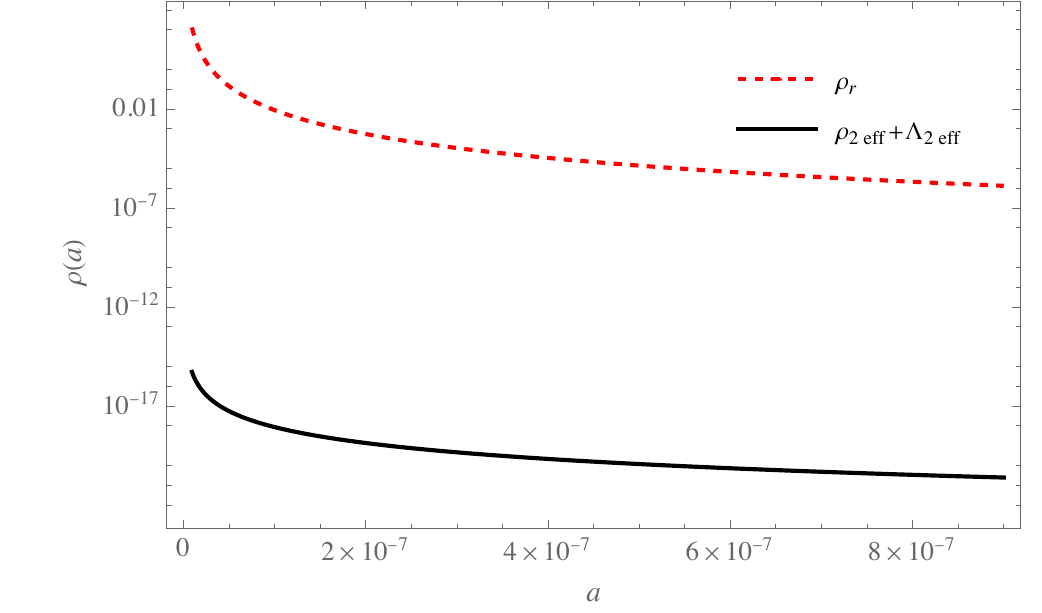}
\caption{Here the parameter values are: $H_0 = 67.4  
\text{km} \cdot\text{s}^{-1}\cdot \text{Mpc}^{-1}$, 
$\Or = 9\times 10^{-5}$, $\Ok = 0$, $\Odr = -1.05$, 
$\Omega_{\alpha_{1},0}=-0.76$, 
and $\Og = 0.06$.
} 
\label{fig4}
\end{figure}

Since we are trying to describe early universes emerging
from this particular case, by assuming $\Og >0$, we can find 
exact expansions around $a=0$. Certainly, we obtain 
two real solutions
\be 
\begin{aligned}
\chi_1 &\simeq  H_0 \left( \frac{\Odr}{\Or} \right),
\\
\chi_2 &\simeq \frac{H_0}{\Og}
\left\lbrace
\frac{(\Og^2 \Or)^{1/3}}{a^{4/3}} - \frac{1}{3} 
\left[
1 + \Og \left( \frac{\Odr}{\Or} \right) 
\right] 
\right.
\\
& \left. + \frac{1}{9} 
\left[
1 - 3\,\Ob \Og - \Og \left( \frac{\Odr}{\Or}\right)
\right. \right.
\\
& \left. \left. - 2\, \Og^2 \left( \frac{\Odr}{\Or}\right)^2
\right] \frac{a^{4/3}}{(\Og^2 \Or)^{1/3}}
\right\rbrace .
\end{aligned}
\ee
With these solutions in hand, and considering~(\ref{Friedmann0}), 
one derives two possible Friedmann-type equations
\beq
H^2 + \frac{k}{a^2} &\approx & \frac{1}{3}\,
\Lambda_{1 \,\text{\tiny eff}},
\label{universe1}
\\
H^2 + \frac{k}{a^2} &\approx & \frac{1}{3}\,
\Lambda_{2\,\text{\tiny eff}} + \frac{1}{3}\,
\rho_{\text{\tiny eff}},
\label{universe2}
\eeq
where we have introduced the effective cosmological 
constants
\beq
\Lambda_{1\,\text{\tiny eff}} &=& 3 H_0^2 \left( 
\frac{\Odr}{\Or} 
\right)^2,
\label{L1}
\\
\Lambda_{2\,\text{\tiny eff}} &=& \frac{H_0^2}{\Og^2}
\left[ 1 - 2\,\Ob\Og - \Og^2 \left( \frac{\Odr}{\Or} 
\right)^2 \right],
\label{L2}
\eeq
and an effective energy density
\be 
\begin{aligned}
\rho_{2\,\text{\tiny eff}} &= \frac{H_0^2}{3\Og^2}
\left\lbrace 
\left( \frac{\Omega_{r,\text{\tiny eff}}}{a^4} \right)^{2/3} 
\right.
\\
& \left. \quad \quad
-\, 2 \left[3 + \Og \left( \frac{\Odr}{\Or}\right)
\right] \left( \frac{\Omega_{r,\text{\tiny eff}}}{a^4} 
\right)^{1/3}
\right\rbrace
\end{aligned}
\label{rho3}
\ee
with $\Omega_{r,\text{\tiny eff}}$ being the same 
as~(\ref{Omega-r-eff}).
Two remarks are worth being mentioning. On the one hand, we 
have a universe driven solely by radiation effects, 
see~(\ref{universe1}) and~(\ref{L1}). On the contrary, the 
other possibility is more interesting and considers the geometry 
due to the presence $\Ob$ and $\Og$ in~(\ref{L2}). 
Concerning this last instance, it is clear how the GHYM-like
term changes the $\rho$ dependence of Friedmann-type
equation to $\rho^{2/3} = \left( \Omega_{r,\text{\tiny eff}}/a^4 
\right)^{2/3}$ in~(\ref{rho3}),
causing the universe to expand fast showing an unconventional 
cosmology.
In this sense, figure~\ref{fig4} compares the sum of the 
effective density and the effective cosmological constant 
with the conventional radiation density. 
The effective density of the modified dark component which 
already includes a radiation-like term remains several orders 
of magnitude below the standard radiation density of the 
reference cosmological model. This suggests that, at least 
for the parameter values considered, the additional 
contributions are screened by the structure of the effective 
density itself and fail to surpass radiation as the dominant 
component during the earliest stages of cosmic evolution.

\begin{figure}[h!]
  \includegraphics[width=1\columnwidth]{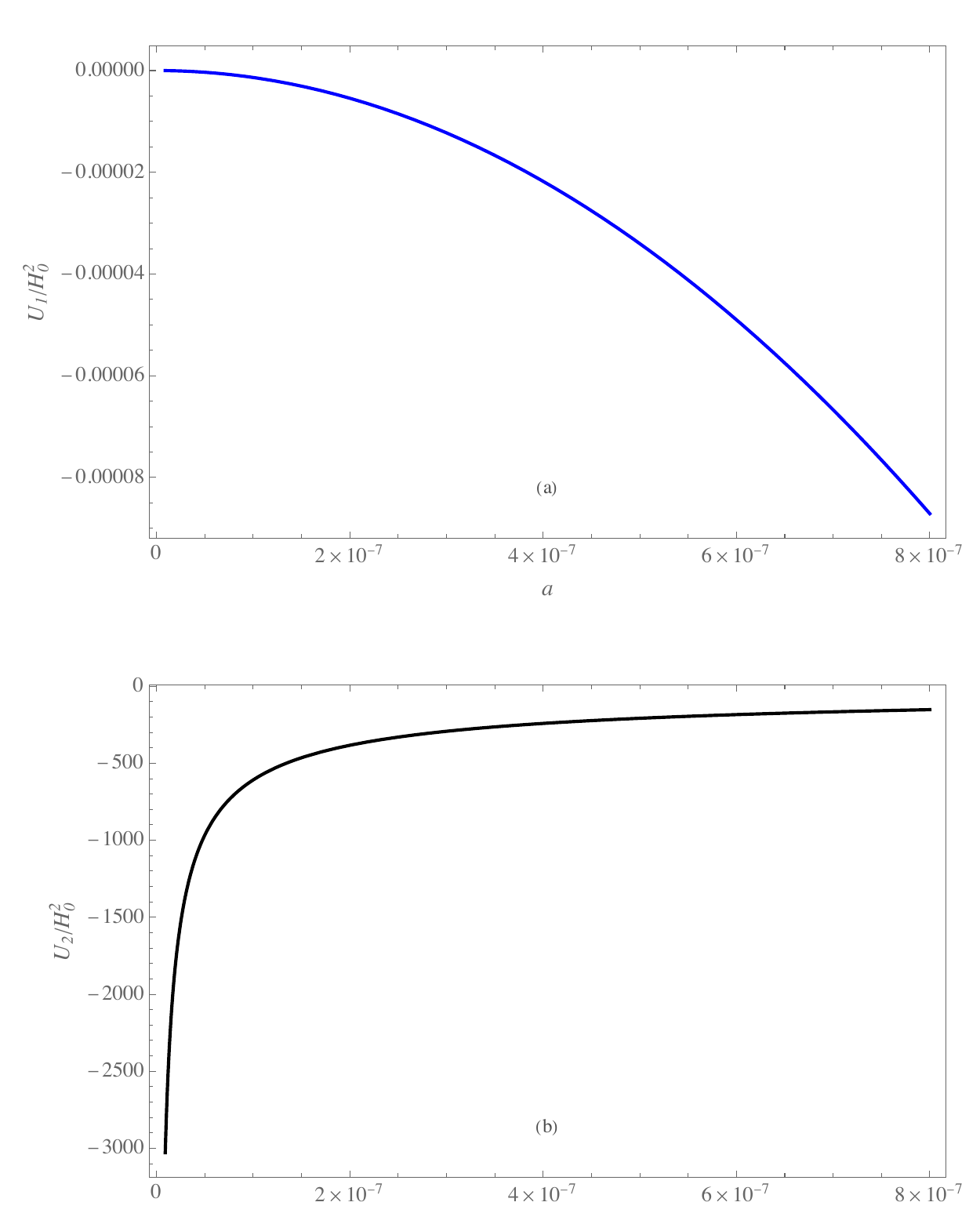}
  \caption{ The parameter values used are: \( H_0 = 67.4 \ \text{km} \cdot\text{s}^{-1}\cdot \text{Mpc}^{-1} \), \( \Omega_{r,0} = 9\times 10^{-5} \), \( \Omega_{k,0} = 0 \), \( \Omega_{dr,0} = -1.05 \), 
 $\Omega_{\alpha_{1},0}=-0.76$, and
 \( \Omega_{\alpha_3,0} = 0.06 \).} 
  \label{fig5}
\end{figure}

To determine the dynamics of these universes we
get the approximated effective potential functions
\beq 
\frac{U_1}{H_0^2} &=& - \Ok - \frac{a^2}{3H_0^2}\,
\Lambda_{1\,\text{\tiny eff}},
\\
\frac{U_2}{H_0^2} &=& - \Ok - \frac{a^2}{3H_0^2}\,
\left[ \Lambda_{2\,\text{\tiny eff}} + 
\rho_{2\,\text{\tiny eff}}(a)
\right],
\eeq
where we have considered~(\ref{L1}),~(\ref{L2}), and~(\ref{rho3}).
In figure~\ref{fig5} we depicted the effective potentials 
$U_1$ and $U_2$, normalized by $H_0^2$, in the early 
universe epoch. The potential $U_1$, proportional to $a^2$, 
remains nearly constant over the range shown, indicating a 
smooth evolution whereas $U_2$  decreases rapidly as $a \to 0$, 
with a milder divergence of the form $U_2(a) \sim -\left(1/a^{2/3}
\right)$.
This behavior defines a dynamical barrier that is weaker 
than that associated with standard radiation, whose 
effective potential diverges as $-1/a^2$. The gentler 
slope of $U_2$ implies that, although the potential becomes 
increasingly negative toward the past, it does not 
significantly restrict the evolution toward $a = 0$.

At this point, one might question whether it is appropriate 
to neglect the matter contribution, given that the model under 
consideration is not linear, in contrast to the standard 
$\Lambda$CDM case. Consequently, it is not evident that the 
baryonic matter component can be safely disregarded at early 
times, as is commonly assumed in the conventional framework. 
However, when performing asymptotic expansions that explicitly 
incorporate this term, the dominant behavior in the early time 
regime remains essentially unchanged. This suggests a screening 
mechanism characteristic of models with higher-order curvature 
corrections. A similar structure can be recognized in some 
of the expressions presented in \cite{Charmousis2002,Kofinas2003}.

\begin{figure}[h!]
\includegraphics[width=1\columnwidth]{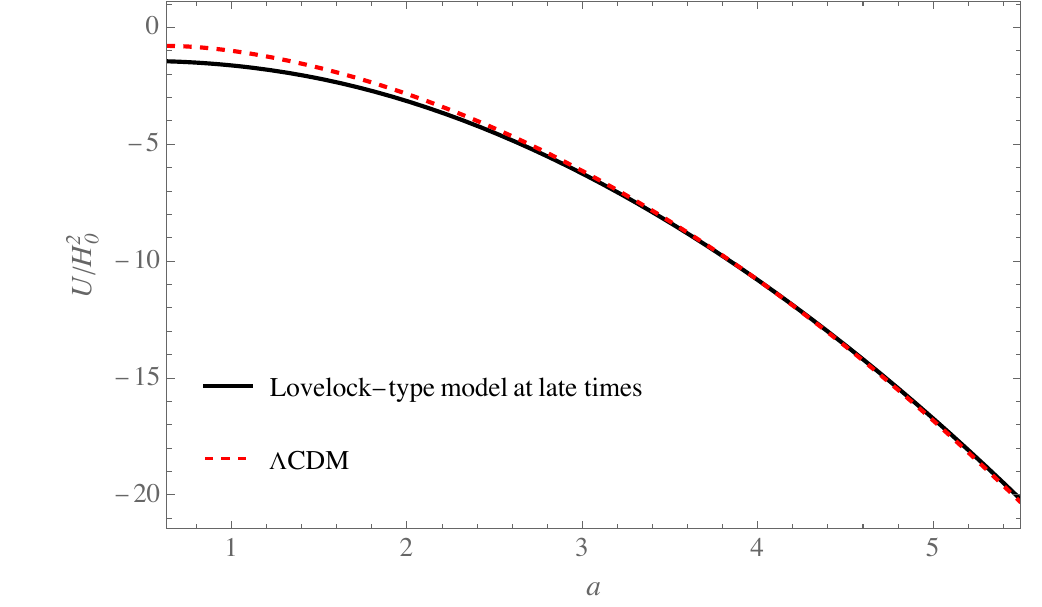}
\caption{ The parameter values used are:   $\Ok = 0$, 
$\Odr =-1.28$, $\Og =-0.61$, $\Ob =1.22$, $\Om =0.33$.
} 
\label{fig_7}
\end{figure}

In this regards, as a valuable approximation to compare with
$\Lambda$CDM scenario, we turn off $\Or$ contribution 
in~(\ref{quartic}). In this regime $\Or$ is effectively 
replaced by $\Om$ along its correspondingly power of scale
factor, $a$, since at late times matter dominates the 
cosmological dynamics. It is also possible to expand the 
roots $\chi_i$ in the large-$a$ regime.Our findings shed 
light on a single expansion that is physically meaningful.
The asymptotic form  of such a solution reads
\be
\begin{aligned}
\chi &=-\frac{1}{2\Omega_{\alpha_3}}\left(1-\sqrt{1-4\Omega_{\alpha_{1,0}}\Omega_{\alpha_{3,0}}}\right)
\\
&- \frac{\Omega_{m,0}}{2\Omega_{\alpha_1}}\left( 1+\frac{1}{\sqrt{1-4\Omega_{\alpha_{1,0}}\Omega_{\alpha_{3,0}}}}
\right)\frac{1}{a^3}.
\end{aligned}
\ee
One can fine-tune the model parameters so that, in the 
large-$a$ regime, the resulting effective potential closely 
matches that of the standard $\Lambda$CDM model. Truncating 
the expansion at order $1/a^3$, one obtains an approximate 
potential whose behavior is illustrated in figure~\ref{fig_7}, 
clearly demonstrating the successful emulation of the standard 
cosmological scenario. 
It is important to note that the values of these parameters 
were chosen so that the model emulates $\Lambda$CDM in the 
late-time regime. They should not necessarily be considered 
suitable for describing the early-time behavior, where the 
neglected terms in the expansion may play a relevant role
and modify their effective values.

\subsection{Dark energy from Lovelock type brane cosmology}

Grounded in the conventional cosmology by enforcing an 
effective FRW evolution dictated by 
\be 
\frac{\dot{a}^2 + k}{a^2} = \frac{1}{3\alpha} \left( \rho 
+ \rho_{\text{\tiny dark}} \right) = : \frac{1}{3\alpha}
\rho_{\text{\tiny eff}} (a),
\label{master3}
\ee
where $\alpha$ is a constant with appropriate units, we wonder 
about the possibility of rearranging (\ref{master}) in this 
way to find out $\rho_{\text{\tiny dark}}$. In fact,
this structure encodes all the additional contributions provided by 
the existence of an extra dimension, as well as the models 
dependent on extrinsic curvature accompanied by $\alpha_1$ and 
$\alpha_3$, to the primitive energy $\rho$. 

In this sense, if (\ref{master}) and~(\ref{master3}) are
in accordance, $\rho_{\text{\tiny dark}}$ results in a root
of a quartic equation. Indeed, by inserting~(\ref{master2})
into~(\ref{master}) we are able to get
\be 
\begin{aligned}
& -\frac{1}{6(3\alpha)^{1/2}} \rho \,(\rho + 
\rho_{\text{\tiny dark}})^{1/2} + \frac{\alpha_1}{6\alpha}
(\rho + \rho_{\text{\tiny dark}}) 
\\
& + \frac{\alpha_2}{(3\alpha)^{3/2}} (\rho 
+ \rho_{\text{\tiny dark}})^{3/2} 
+ \frac{\alpha_3}{(3\alpha)^2} (\rho 
+ \rho_{\text{\tiny dark}})^2
\\
& \qquad \qquad \qquad \qquad \qquad \,\,
+ \frac{\omega}{a^4} = 0.
\end{aligned}
\ee
Now we introduce $\mathcal{Z}:= \left[ \rho + 
\rho_{\text{\tiny dark}} /(3\alpha) \right]^{1/2}$ and 
reorganize this equation to obtain
\be 
\begin{aligned}
& \Og \,\mathcal{Z}^4 + H_0\,\mathcal{Z}^3
+ \Ob\,\mathcal{Z}^2
\\
& 
+ \left[ \OL - \left( \frac{\Om}{a^3} + \frac{\Or}{a^4} \right)
\right] H_0^3 \,\mathcal{Z} + \frac{\Odr H_0^4}{a^4} = 0,
\end{aligned}
\label{quartic2}
\ee
where we have used the energy densities~(\ref{densities}). 
This is the same equation we already obtained 
in~(\ref{quartic}).
Depending on the values chosen for the $\Omega_I$s, we could 
have real solutions to analyze, as the discriminant of the 
quartic equation dictates,~\cite{Prodanov2021}. In view of the
structures~(\ref{structures}) the solutions for $\mathcal{Z}$ 
are similar to~(\ref{chi0}), so the general form for 
$\rho_{\text{\tiny dark}}$ is given by
\begin{widetext}
\be 
\begin{aligned}
\rho_{\text{\tiny dark}} &= - \rho(a) + 
\frac{3\alpha H_0^2}{16\Og^2}
\left\lbrace
\sgn (\Og) \pm \sqrt{u + \frac{4}{3} 
\left[ 
\frac{g(\Omega_I,a)}{h(\Omega_I,a)} + h(\Omega_I,a)
\right]
\Og}
\right.
\\
&\qquad \qquad \qquad \qquad \qquad \qquad \quad \pm \left. 
\sqrt{
2\,u - \frac{4}{3}\left[ \frac{g(\Omega_I,a)}{h(\Omega_I,a)} 
+ h(\Omega_I,a)\right]\Og 
+ \frac{\sgn (\Og)\,2\, s(\Omega_I,a)}{\sqrt{u 
+ \frac{4}{3} 
\left[ 
\frac{g(\Omega_I,a)}{h(\Omega_I,a)} + h(\Omega_I,a)
\right]
\Og}
}
}
\right\rbrace^2,
\end{aligned}
\label{rho-dark}
\ee
\end{widetext}
where we must keep in mind the combinations of the signs, 
providing four possible roots. 
This accommodates all the extra contributions, physical 
and geometrical, provided by the correction terms $\Ob$, $\Og$, 
and the extra dimension, $\Odr$, to the primitive energy 
density $\rho$. In a sense, the evolution of these types 
of universes is dictated by an effective energy density 
$\rho_{\text{\tiny eff}} = \rho_{\text{\tiny dark}} + \rho(a)$ 
and not solely by the primitive energy density $\rho$.
In passing, as already mentioned, the real solutions are 
provided by an appropriate choice of the $\Omega_I$s according 
to the rules established by the discriminant of the quartic equation~(\ref{quartic2}),~\cite{Prodanov2021}.

A couple of remarks are in order. It is worthwhile to observe
that~(\ref{rho-dark}) must be accompanied by the normalization condition~(\ref{normalization}). Also,~(\ref{rho-dark}) 
guarantees the definite positivity of the total energy
density $\rho + \rho_{\text{\tiny dark}}$. 
To conclude this subsection, we must recognize that the study of~(\ref{rho-dark}) offers another perspective for analyzing the dynamics of emerging universes in this model. Indeed, from this result, we can determine an effective pressure and an effective 
equation of state as more appropriate expressions for 
understanding cosmological effects, after fine-tuning, 
to compare them with observational data. 

\section{Concluding remarks}
\label{sec5}
In this work we have developed systematically a cosmological
framework for geodetic brane gravity enhanced with
Lovelock-type invariants defined on the world volume swept out
by an extended object. Being a purely geometric theory, for a
4-dimensional brane-like universe the GHY- and GHYM-type
invariants that are allowed to come into play alone provide
interesting cosmological implications.
This combined model leads to a second-order equation of motion
for the field variables, ensuring that no additional non-physical
degrees of freedom appear, and preserves reparametrization invariance
of the world volume. For a homogeneous and isotropic embedding
in a 5-dimensional Minkowski spacetime, such invariance under reparameterizations leads to the appearance of the integration
constant $\omega$, the fingerprint of the extra dimension, which
parametrizes the deviation from the usual
Einstein gravity, DGP, or certain regimen
of the GB brane cosmologies by analyzing a set of possible Friedmann-type equations. Indeed, the model allows scenarios akin
to some accelerating brane cosmological models. It should be noted,
however, that these scenarios correspond to quite different situations.
In a manner, what we have given classically is by no means a
complete analysis of the model. Our aim is to highlight that this special type of geometric
invariants can mimic, under certain conditions, the accelerated
behavior of our universe.
The GHY- and GHYM-type terms become relevant at late times, providing
a mechanism for cosmic acceleration without the need to introduce
additional exotic components. In this sense, dark energy appears as
a purely geometric contribution, arising from the combination of
the extrinsic terms together with the integration constant $\omega$,
what is also evident in this framework, by rewriting the Friedmann-type equations. Indeed, effective energy driving evolution arises from
the combination of ordinary matter and a companion density one
emerging from the embedding geometry which is evident
in~(\ref{rho-dark}). 

In the early regime, dominated by radiation, the cubic term in the extrinsic curvature produces non-standard dependences that modify the expansion compared to conventional cosmology. With a suitable choice of parameters, the model closely reproduces the dynamics of \(\Lambda\)CDM at late times, while at the same time providing controlled deviations in the early universe.  

The generalized normalization condition imposes relations among the dimensionless densities of the model, and the reality of the solutions to the quartic equation determines which branches lead to physically viable evolutions. Within these domains, Lovelock-type brane cosmology may be regarded as a scenario that connects known theories and allows the exploration of new dynamics.

To sum up, Lovelock-type brane cosmology provides a coherent and
geometrically motivated framework, second order in nature, which not
only reproduces the standard results in the appropriate limits,
but also offers alternative and unified mechanisms to explain cosmic acceleration and the emergence of an effective dark sector.  
Along with, an interesting issue to be addressed lies in the
quantum implications of the model, following the line of reasoning
given in~\cite{Rojas2014}, since the model inherits the property of
being an affine in accelerations theory~\cite{Rojas2009}, which
would allow us to know whether the embryonic epoch,~\cite{Davidson:1999fb,Rojas2009,Rojas2014}, characteristic of this type of models still persists or has changed substantially.

\acknowledgments

RC acknowledges support by COFAA-IPN, Est\'\i mulos al Desempe\~no 
de los Investigadores (EDI)-IPN and SIP-IPN Grants No.~20241624 and 20251228. ER thanks ProdeP-M\'exico, CA-UV-320: \'Algebra, 
Geometr\'\i a y Gravitaci\'on.
GC acknowledges support from a Postdoctoral Fellowship
by Estancias Posdoctorales por México 2023(1)-SECIHTI. This work
was partially supported by Sistema Nacional de Investigadoras e 
Investigadores, México.



\begin{thebibliography}{}

\bibitem{Riess:1998}
A. G. Riess et al. (Supernova Search Team), Astron. J. \textbf{116}, 1009 (1998).

\bibitem{Perlmutter:1999}
S. Perlmutter et al. (Supernova Cosmology Project), Astrophys. J. \textbf{517}, 565 (1999),

\bibitem{Wetterich:1987fm}
Wetterich, Nucl. Phys. B \textbf{302}, 668 (1988).

\bibitem{Zlatev:1998tr}
I.~Zlatev, L.~M.~Wang and P.~J.~Steinhardt,
Phys. Rev. Lett. \textbf{82}, 896-899 (1999).


\bibitem{Kading:2023hdb}
C.~K{\"a}ding,
Astronomy \textbf{2}, no.2, 128-140 (2023).

\bibitem{Burrage:2018zuj}
C.~Burrage, E.~J.~Copeland, C.~K{\"a}ding and P.~Millington,
Phys. Rev. D \textbf{99}, no.4, 043539 (2019)

\bibitem{Chiba:2000}
 T. Chiba, T. Okabe, and M. Yamaguchi, Phys. Rev. D \textbf{62}, 023511 (2000).
 
\bibitem{Armendariz-Picon:2000} 
C. Armendariz-Picon, V. F. Mukhanov, and P. J. Steinhardt, Phys. Rev. Lett. \textbf{85}, 4438
(2000)

\bibitem{Armendariz-Picon:2001} 
C. Armendariz-Picon, V. F. Mukhanov, and P. J. Steinhardt, Phys. Rev. D \textbf{63}, 103510
(2001)

\bibitem{Melchiorri:2003} 
A. Melchiorri, L. Mersini-Houghton, C. J. Odman, and M. Trodden, Phys. Rev. D \textbf{68}, 043509
(2003)

\bibitem{Chimento:2003} 
L. P. Chimento and A. Feinstein, Mod. Phys. Lett. A \textbf{19}, 761 (2004).

\bibitem{Chimento:2004}
L. P. Chimento, Phys. Rev. D \textbf{69}, 123517 (2004).

\bibitem{Garriga:1999}
J. Garriga and V. F. Mukhanov, Phys. Lett. B \textbf{458}, 219 (1999).

\bibitem{Armendariz-Picon:1999}
C. Armendariz-Picon, T. Damour, and V. F. Mukhanov, Phys. Lett. B \textbf{458}, 209 (1999).

\bibitem{Putter:2007}
R. de Putter and E. V. Linder, Astropart. Phys. \textbf{28}, 263 (2007).

\bibitem{Gao:2010}
X.-T. Gao and R.-J. Yang, Phys. Lett. B \textbf{687}, 99 (2010).

\bibitem{Nicolis:2009}
A. Nicolis, R. Rattazzi, and E. Trincherini, Phys. Rev. D \textbf{79}, 064036 (2009).

\bibitem{Rham:2010}
C. de Rham and A. J. Tolley, JCAP \textbf{05}, 015 (2010).

\bibitem{Goon:2011}
G. L. Goon, K. Hinterbichler, and M. Trodden, Phys. Rev. D \textbf{83}, 085015 (2011).

\bibitem{Goon:2011J}
G. Goon, K. Hinterbichler, and M. Trodden, JCAP \textbf{07}, 017 (2011).

\bibitem{Scherrer:2004}
R. J. Scherrer, Phys. Rev. Lett. \textbf{93}, 011301 (2004).

\bibitem{Deffayet:2010}
C. Deffayet, O. Pujolas, I. Sawicki, and A. Vikman, JCAP \textbf{10}, 026 (2010).

\bibitem{Pujolas:2011}
O. Pujolas, I. Sawicki, and A. Vikman, JHEP \textbf{11}, 156 (2011).

\bibitem{Kimura:2011}
R. Kimura and K. Yamamoto, JCAP \textbf{04}, 025 (2011).

\bibitem{Maity:2013}
D. Maity, Phys. Lett. B \textbf{720}, 389 (2013).

\bibitem{Capozziello:2002rd}
S.~Capozziello,
Int. J. Mod. Phys. D \textbf{11}, 483-492 (2002).

\bibitem{Capozziello:2003gx}
S.~Capozziello, V.~F.~Cardone, S.~Carloni and A.~Troisi,
Int. J. Mod. Phys. D \textbf{12}, 1969-1982 (2003).

\bibitem{Amendola:1999qq}
L.~Amendola,
Phys. Rev. D \textbf{60}, 043501 (1999).

\bibitem{Bartolo:1999sq}
N.~Bartolo and M.~Pietroni,
Phys. Rev. D \textbf{61}, 023518 (2000).

\bibitem{Carroll:2004de}
S.~M.~Carroll, A.~De Felice, V.~Duvvuri, D.~A.~Easson, M.~Trodden and M.~S.~Turner,
Phys. Rev. D \textbf{71}, 063513 (2005).

\bibitem{Nojiri:2005vv}
S.~Nojiri, S.~D.~Odintsov and M.~Sasaki,
Phys. Rev. D \textbf{71}, 123509 (2005).

\bibitem{Maartens:2010ar}
R.~Maartens and K.~Koyama,
Living Rev. Rel. \textbf{13}, 5 (2010).

\bibitem{Stern2023}
A. Stern and Chuang Xu, 
Phys. Rev. D 107, 024001 (2023).

\bibitem{RT1975}
T.~Regge and C.~Teitelboim, General Relativity \`a la string: A
progress reprot, in \textit{Proceedings of the First Marcel Grossman 
Meeting}, ed. R. Ruffini (North-Holland, Amsterdam, 1977), p. 77;
arXiv: 1612.05256 [hep-th]. 

\bibitem{Rubakov1983}
V. A. Rubakov and M. E. Shaposhnikov, Phys. Lett. B \textbf{125},
136 (1983).

\bibitem{Rojas2012}
M.~Cruz and E.~Rojas, Class. Quant. Grav. \textbf{30}, 115012 (2013).

\bibitem{Rojas2014}
R. Cordero, M. Cruz, A. Molgado and E. Rojas, Gen. Rel. Grav.
\textbf{46}, 1761 (2014).

\bibitem{Rojas2015}
A.~Montiel, N.~Bret\'on, R.~Cordero and E.~Rojas,
Phys. Rev. D \textbf{92}, 024042 (2015).

\bibitem{Rojas2013}
M. Cruz and E. Rojas, Class. Quant. Grav. \textbf{30}, 
115012 (2013).

\bibitem{Rojas2016}
N. Bagatella-Flores, C. Campuzano, M. Cruz and E. Rojas,
Class. Quant. Grav. \textbf{33}, 245012 (2016).

\bibitem{Rojas2019}
E. Rojas, Class. Quant. Grav. \textbf{36}, 185006 (2019)

\bibitem{Rojas2025}
E. Rojas and G. Cruz, Phys. Lett. B \textbf{866},  139504 (2025)

\bibitem{Rojas2024}
E. Rojas, G. Cruz and J.C. Natividad, Int. J. Mod. Phys. A 
\textbf{39}, 2450069 (2024)

\bibitem{DGP} 
G.~R.~Dvali, G.~Gabadadze and M.~Porrati, Phys. Lett. B 
\textbf{484}, 112 (2000); \textbf{485}, 208 (2000).

\bibitem{Charmousis2002}
Charmousis, C., and Dufaux, J. F.  Class. Quantum Grav. 
\textbf{19} 4671, (2002).

\bibitem{Kofinas2003}
G. Kofinas, R. Maartens and E. Papantonopoulos, JHEP \textbf{10},
066 (2003).

\bibitem{Myers1987} 
R. C. Myers, Phys. Rev. D \textbf{36}, 392 (1987).

\bibitem{Davis2003} 
S. C. Davis, Phys. Rev. D \textbf{67}, 024030 (2003).

\bibitem{Davidson1999} 
A. Davidson, Class. Quant. Grav. \textbf{16}, 653 (1999).

\bibitem{Davidson2003} 
D. Karasik and A Davidson, Phys. Rev. D \textbf{67}, 
064012 (2003).

\bibitem{Davidson2006} 
A. Davidson and I. Gurwich, Phys. Rev. D {\bf 74}, 044023 (2006).

\bibitem{Spivak1970}
M. Spivak, \textit{A comprehensive introduction to differential 
geometry}, Vol. 4, (Publish or Perish, Boston, MA, 1970) (1979).

\bibitem{Defo1995}
R. Capovilla and J. Guven, Phys. Rev. D \textbf{51}, 6736 (1995) 


\bibitem{Noether2000}
G. Arreaga, R. Capovilla and J. Guven, Annals Phys. \textbf{279},
126-158 (2000)

\bibitem{Davidson2001}
A. Davidson, D. Karasik and Y. Lederer, Cold dark matter from dark 
energy, arXiv:gr-qc/0111107.

\bibitem{Paston}
S. A. Paston and A. A. Sheykin, Eur. Phys. J. C \textbf{78}, 1 (2018).

\bibitem{Prodanov2021}
E. M. Prodanov, Int. J. Appl. Comput. Math. \textbf{7}, 218
(2021).

\bibitem{Davidson:1999fb}
A.~Davidson, D.~Karasik and Y.~Lederer, Class.
 Quant. Grav. \textbf{16} 1349 (1999).

\bibitem{Rojas2009}
R. Cordero, A. Molgado and E. Rojas, Phys. Rev. D \textbf{79}, 
024024 (2009)

\end{thebibliography}
\end{document}